\documentclass[showpacs,amsfonts,amsmath,amssymb]{revtex4}
\usepackage{graphicx}

\def\[{\left[}
\def\]{\right]}
\def\({\left(}
\def\){\right)}

\begin{document}

\title{The general features of Bianchi-I cosmological models in Lovelock gravity}

\author{S.A. Pavluchenko}
\affiliation{Special Astrophysical Observatory, Russian Academy of
Sciences, Nizhnij Arkhyz, 369167 Russia}

\begin{abstract}
We derived equations of motion corresponding to Bianchi-I
cosmological models in the Lovelock gravity. Equations derived in
the general case, without any specific ansatz for any number of
spatial dimensions and any order of the Lovelock correction. We
also analyzed the equations of motion solely taking into account
the highest-order correction and described the drastic difference
between the cases with odd and even numbers of spatial dimensions.
For power-law ansatz we derived conditions for Kasner and
generalized Milne regimes for the model considered. Finally, we
discuss the possible influence of matter in the form of perfect
fluid on the solutions obtained.
\end{abstract}

%04.20.Dw    Singularities and cosmic censorship
%04.20.Fy    Canonical formalism, Lagrangians, and variational principles
%04.20.Jb    Exact solutions
%04.25.dc    Numerical studies of critical behavior, singularities, and cosmic censorship
%04.50.-h    Higher-dimensional gravity and other theories of gravity
%04.50.Gh    Higher-dimensional black holes, black strings, and related objects
%04.50.Kd    Modified theories of gravity
%98.80.-k    Cosmology


\maketitle

\section{Introduction}

The idea that our space-time has more than four dimensions is not
new in science. And since its first introduction by
Kaluza~\cite{Kaluza} and Klein~\cite{Klein} it went a long way
with different inspirations. Nowadays, the interest to such
theories is mostly motivated by the superstrings and supergravity
theories (see, e.g., \cite{super}). The presence of extra
dimensions makes it possible for superstrings theory to exist, but
it leaves us with a gap between the number of dimensions the
superstrings exist in, and our observable four-dimensional
space-time. One of the possible ways to solve this problem is to
make additional dimensions very small on cosmological scales by
means of the cosmological theory.

It is well-known~\cite{etensor1, etensor2, etensor3} that the
Einstein tensor is, in any dimension, the only symmetric and
conserved tensor depending only on the metric and its first and
second derivatives (with a linear dependence on second
derivatives). If one drops the condition of linear dependence on
second derivatives, one can obtain the most general tensor which
satisfies other mentioned conditions -- Lovelock
tensor~\cite{Lovelock}. The corresponding Lagrangian (Lovelock
Lagrangian), which generalizes the Einstein-Hilbert one, can be
obtained by the same way -- the Euler variation with respect to
the metric. The Lovelock tensor differs from the Einstein tensor
only if space-time has more than four dimensions, so the Lovelock
gravity can be considered as a natural generalization of General
Relativity in a higher-dimensional space-time. It contains
quadratic and higher-order corrections in the Lagrangian,
depending on the number of spatial dimensions.

%Theories of gravity with higher-order corrections have plenty of
%implementations in modern cosmology and physics, like
%$R^2$-inflation~\cite{r2} or study of black-hole-like~\cite{bh} or
%wormhole-like~\cite{wh} solutions in Lovelock gravity.

In the cosmological context, the Lovelock gravity has as well been
intensively studied (see, e.g.,~\cite{d4, deruelle1, raz1, raz2,
deruelle2}). In our paper we study the Bianchi-I-type cosmological
model in Lovelock gravity. As we pointed out above, the
anisotropic cosmological models can possibly solve the issue with
small extra dimensions. But to construct a self-consistent model,
freed from singularities and other obstacles, one needs to fully
understand the dynamics of such a system.

%We chose Bianchi-I as a ``toy-model''; the models with different
%$L_n$ seems to share some common features, so the models with a
%more complicated topology can share some as well.

We decided to investigate Bianchi-I cosmological models. These are
models with a simplest geometry and they can be used to understand
common features of the model considered. The case with most
general kind of geometry is more interesting, of course, but is
much more complicated as well. And knowing some features from the
Bianchi-I model analysis could be a great asset while studying
more general cases. Not to mention, exploring Bianchi-I models is
profitable by itself -- indeed, according to nowadays cosmological
point of view, our Universe is spatially flat, so as Bianchi-I
models are, while general case, which usually involves products of
Lie algebras, is not.

As a part of the analysis we demonstrate the difference between
the cases with odd and even numbers of spatial dimensions. In our
previous paper dedicated to the numerical investigation of the
Gauss-Bonnet cosmologies~\cite{mpla} we noticed that there is a
clear difference in the dynamics of (4+1)- and (5+1)-dimensional
models. Now we demonstrate that this difference is generic for all
orders of Lovelock correction and for any number of dimensions.

Finally we take the matter in form of the perfect fluid into
analysis. This decision is also motivated by our previous studies.
In~\cite{new} we discovered the existence of ``special solution''
for pressureless fluid in (4+1)-dimensional Gauss-Bonnet
cosmology; in this paper, among other things, we demonstrate that
this is the common feature of models with even number of spatial
dimensions.

The structure of a manuscript is as follows: first, we derive the
Lagrangian for the model with only one highest possible $n$ taken
into account as well as equations of motion. Then, we consider
cases with even and odd numbers of spatial dimensions separately,
demonstrate the difference in the structure of the equations and,
as a consequence, a difference in their dynamics. In section IV we
apply so-called power-law {\it anzatz} to our equations and obtain
conditions for Kasner and generalized Milne regimes; conditions
obtained generalize known (for low $n$) ones. Then we take into
account matter in form of a perfect fluid and investigate the
influence of it onto the dynamics of the system. In Discussion we
discuss the topics relevant to the results obtained and finally in
the Conclusion sum up all the results.

\section{Equations of motion}

Lovelock invariants have the form~\cite{Lovelock}

\begin{equation}
L_n = \frac{1}{2^n}\delta^{i_1 i_2 \dots i_{2n}}_{j_1 j_2 \dots
j_{2n}} R^{j_1 j_2}_{i_1 i_2}
 \dots R^{j_{2n-1} j_{2n}}_{i_{2n-1} i_{2n}}, \label{lov_lagr}
\end{equation}

\noindent where $\delta^{i_1 i_2 \dots i_{2n}}_{j_1 j_2 \dots
j_{2n}}$ is the generalized Kronecker delta of the order $2n$.
%Since we are interested in the behavior near the initial
%singularity, we can take into account only the highest order
%correction, so the Lagrangian density takes a form
The Lagrangian density have a form

\begin{equation}
{\cal L}= \sqrt{-g} \sum_n \alpha_n L_n, \label{lagr}
\end{equation}

\noindent where $g$ is the determinant of metric tensor,
$\alpha_n$ is a constant of the order of Planck length in $n$
dimensions and summation over all $n$ in consideration is assumed.
In our analysis we take into account only one highest possible
correction -- we are interested in the behavior in the vicinity of
the cosmological singularity. In that regime higher curvature
corrections are dominating over lower ones so one can consider
highest possible correction only. In that case (\ref{lagr})
contains only one term with highest possible $n$:

\begin{equation}
{\cal L}= \sqrt{-g} L_n, \label{lagr1}
\end{equation}

\noindent since we have only one $n$, we can renormalize $L_n$ to
set $\alpha_n$ to unity. As we mentioned in Introduction, we work
with Bianchi-I-type metric in $(D+1)$-dimensional space-time:

$$
ds^2 = {\rm diag} (-1,\, a_1^2(t),\, a_2^2(t),\,\ldots\,
a_D^2(t)),
$$

\noindent where
%$N(t)$ is a lapse function and
$a_i(t)$ is a scale factor, corresponding to $i$-th spatial
dimension. From this metric one can build the following non-zero
Riemann tensor components:

\begin{equation}
R_{0i0i} = -a_i(t) \ddot a_i(t), \quad R_{ijij} = a_i(t) \dot
a_i(t) a_j(t) \dot a_j(t), \label{riemann1}
\end{equation}

\noindent where Latin indices correspond to the spatial
coordinates, the dot above corresponds to a derivation with
respect to time and standard relations between indices for Riemann
tensor are kept in mind (see, e.g.,~\cite{textbooks}). By raising
indices in Eq.\,(\ref{riemann1}) one can get

\begin{equation}
R^{0i}_{0i} = \frac{\ddot a_i(t)}{a_i(t)}, \quad R^{ij}_{ij} =
\frac{\dot a_i(t)
 \dot a_j(t)}{
a_i(t) a_j(t)}. \label{riemann2}
\end{equation}

Since we are working in a flat space-time, it is convenient to
rewrite Eq.\,(\ref{riemann2}) in terms of the Hubble parameter
$H_i \equiv \dot a_i(t)/ a_i(t)$:

\begin{equation}
R^{0i}_{0i} = \dot H_i + H_i^2, \quad \quad R^{ij}_{ij} = H_i H_j.
\label{riemann3}
\end{equation}

Now we can substitute (\ref{riemann3}) into (\ref{lov_lagr}),
imply the properties of the generalized Kronecker delta and obtain
the expression for $n$-th order Lovelock invariant for
Bianchi-I-type metrics:

%$$
%\begin{array}{l}
%L_n^D = (2n-3) \sum\limits^D_{i=1} \left[ (\dot H_i + H_i^2) \sum\limits_{i\ne\{j_1 > j_2 > \dots j_{2n-2}\}}  H_{j_1} H_{j_2} \dots H_{j_{2n-2}} \right] + \\ \qquad + (2n-1)\sum\limits_{k_1 > k_2 > \dots k_{2n}}
%H_{k_1} H_{k_2} \dots H_{k_{2n}},
%\end{array} \eqno(5)
%$$

\begin{equation}
\begin{array}{l}
L_n^D = 2^{-n}\left((2n-3)!! \sum\limits^D_{i=1} \left[ (\dot H_i
+ H_i^2)
\sum\limits_{\substack{\{j_1,\,j_2,\,\dots\,j_{2n-2}\}\ne i\\
j_1
> j_2 > \dots j_{2n-2}}}  H_{j_1} H_{j_2} \dots H_{j_{2n-2}}
\right]\right. +
\\ \left.\qquad + (2n-1)!!\sum\limits_{k_1 > k_2 > \dots k_{2n}} H_{k_1} H_{k_2} \dots
H_{k_{2n}}\right),
\end{array} \label{big_lagr1}
\end{equation}

\noindent where $(2n-1)!!$ is bifactorial of $(2n-1)$: $(2n-1)!! =
(2n-1)\times (2n-3) \times\cdots$. There $D$ is a number of
spatial dimensions and $D \geqslant 2n$.
%Now if we are interested
%only in the highest-order corrections, we only consider the
%highest-$n$ contribution.
Since we deal with only one (highest possible) $n$, we can neglect
the $1/2^n$ multiplier as well as cancel $(2n-3)!!$ with a
reduction of $(2n-1)!!$ and Eq.\,(\ref{big_lagr1}) become

\begin{equation}
\begin{array}{l}
L_n^D = \sum\limits^D_{i=1} \left[ (\dot H_i + H_i^2)
\sum\limits_{\substack{\{j_1,\,j_2,\,\dots\,j_{2n-2}\}\ne i\\
j_1
> j_2 > \dots j_{2n-2}}}  H_{j_1} H_{j_2} \dots H_{j_{2n-2}}
\right] +
\\ \qquad + (2n-1)\sum\limits_{k_1 > k_2 > \dots k_{2n}} H_{k_1} H_{k_2} \dots
H_{k_{2n}}.
\end{array} \label{big_lagr2}
\end{equation}

%If one leave only leading terms (highest possible curvature order)
%then $D=2n$ for even and $D=2n+1$ for odd number of spatial
%dimensions.
The first term in Eqs.\,(\ref{big_lagr1}) and (\ref{big_lagr2})
comes from the products of a $R^{0i}_{0i} R^{jk}_{jk} \dots$ kind
and the second one -- from the products like $R^{ij}_{ij}
R^{kl}_{kl} \dots$. Since the same index can not appear twice, the
products of a $R^{0i}_{0i} R^{0j}_{0j} \dots$ kind are impossible.
%Also one can
%easily notice Eqs.\,(\ref{big_lagr1}) and (\ref{big_lagr2}) are
%correct only for $n\geqslant 2$ -- since for $L_1$ (which is just
%$R$) there are no products of mentioned above types.

Then one can substitute Eq.\,(\ref{big_lagr2}) into
Eq.\,(\ref{lagr1}) and infer the equations of motion via varying;
thus, the constraint equation takes a form

\begin{equation}
(2n-1) \sum\limits_{k_1 > k_2 > \dots k_{2n}} H_{k_1} H_{k_2}
\dots H_{k_{2n}} = 0; \label{constr_gen}
\end{equation}

\noindent the dynamical equation corresponding to $i$-th spatial
coordinate becomes

\begin{equation}
\begin{array}{l}
\sum\limits^D_{\substack{m=1 \\ m\ne i}} \left[ (\dot H_m + H_m^2)
\sum\limits_{\substack{\{j_1,\,j_2,\,\dots\,j_{2n-2}\}\ne\{i, m\}\\
j_1
> j_2 > \dots j_{2n-2}}}  H_{j_1} H_{j_2} \dots H_{j_{2n-2}}
\right] +
\\ \qquad + (2n-1)\sum\limits_{\substack{\{ k_1, k_2, \dots k_{2n}\}\ne i\\
k_1 > k_2 > \dots k_{2n}}} H_{k_1} H_{k_2} \dots
H_{k_{2n}} = 0.
\end{array} \label{dyn_gen}
\end{equation}

\section{The difference between the $D=2n$ and $D=2n+1$ cases}

Eqs.\,(\ref{constr_gen}) and (\ref{dyn_gen}) might look
complicated but in fact they are quite simple; to illustrate it
let us rewrite them for odd and even numbers of spatial
dimensions. For this one can notice, that any given $n$ gives a
maximal-order correction for an odd $D=2n+1$ and even $D=2n$
number of spatial dimensions. So for odd $(D=2n+1)$ number of
spatial dimensions, the constraint equation takes a form

\begin{equation}\label{constr_odd}
\sum\limits_{k_1 > k_2 > \dots k_{2n}} H_{k_1} H_{k_2} \dots
H_{k_{2n}} = 0;
\end{equation}

\noindent this is just the sum of all possible multiplications put
in order with $2n$ multipliers constructed from Hubble parameters.
One can see there are only $(2n+1)$ terms in this sum, and we
omitted the $(2n+1)$ multiplier in front of sum in
(\ref{constr_gen}) since it is always non-zero. The dynamical
equation corresponding to $i$-th spatial coordinate become

\begin{equation}\label{dyn_odd}
\begin{array}{l}
\sum\limits^D_{\substack{m=1 \\ m\ne i}} \left[ (\dot H_m + H_m^2)
\sum\limits_{\substack{\{j_1,\,j_2,\,\dots\,j_{2n-2}\}\ne\{i, m\}\\
j_1
> j_2 > \dots j_{2n-2}}}  H_{j_1} H_{j_2} \dots H_{j_{2n-2}}
\right] +
\\ \qquad + (2n-1)~ H_{k_1} H_{k_2} \dots H_{k_{2n}} = 0.
\end{array}
\end{equation}

\noindent For the case considered, the second term in
Eq.\,(\ref{dyn_gen}) has only one possible combination -- indeed,
we have $(2n+1)$ spatial dimensions, one of them is ``forbidden''
(since this equation corresponds to the $i$-th coordinate) and the
remaining $2n$ coordinates have only one multiplication
combination with $2n$ multipliers.

For an even number of spatial dimensions $(D=2n)$ we have a
constraint equation in the following form:

\begin{equation}\label{constr_even}
H_{k_1} H_{k_2} \dots  H_{k_{2n}} = 0.
\end{equation}

\noindent The same with the Eq.\,(\ref{constr_odd}) reasoning, we
have only one multiplication with $2n$ multipliers constructed
from Hubble parameters for this case. And this multiplication is
equal to zero. This leaves us the first drastic difference between
the cases with odd and even numbers of spatial dimensions -- the
latter should have one of Hubble parameters always equal to zero.
The dynamical equations for the odd-number case:

\begin{equation}\label{dyn_even}
%\begin{array}{l}
\sum\limits^D_{\substack{m=1 \\ m\ne i}}  (\dot H_m +
H_m^2)~H_{j_1} H_{j_2} \dots H_{j_{2n-2}} = 0.
%\end{array}
\end{equation}

\noindent One can clearly see the difference between
Eqs.\,(\ref{dyn_gen}) and (\ref{dyn_odd}): the last term vanished
due to the absence of multiplications of this type and the r.h.s.
of the first term is a multiplication with $2n-2$ terms: one index
is ``forbidden'' because it's an $i$-th dynamical equation,
another one comes into $(\dot H_m + H_m^2)$ part, consequently we
are left exactly with $2n-2$ ``free'' Hubble parameters which form
only one put in order multiplication with $2n-2$ multipliers.

Therefore one can clearly see the difference between the cases
with odd and even number of spatial dimensions -- they differ both
in the constraint equation as well as in dynamical ones.
In~\cite{new} (and previously in~\cite{mpla}) we showed the
difference between the $(4+1)$ and $(5+1)$ Gauss-Bonnet-dominated
($L_2$) Universe; now we see that this difference is global and
that it is applied to all orders of Lovelock gravity.

\section{Power-law anzatz}

Since the Kasner's paper~\cite{kasner}, the power-law {\it anzatz}
for Bianchi-I model is widely used for various reasons. It implies
the scale factors to be written as $a_i(t) = t^{p_i}$, so the
Hubble parameters would be $H_i = p_i / t$ and the derivation of
Hubble parameters with respect to time would have the form $\dot
H_i = - p_i / t^2$. Let us rewrite the equations of motion using
this {\it anzatz} and obtain some important results. Thus,
constraint equation of our system takes a form

\begin{equation}
%\frac{(2n-1)}{t^{2n}}
\sum\limits_{k_1 > k_2 > \dots k_{2n}} p_{k_1} p_{k_2} \dots
p_{k_{2n}} = 0; \label{constr_gen_p}
\end{equation}

\noindent we omit the $(2n-1)/t^{2n}$ multiplier since only one
$n$ (the highest possible) is taken into account and we deal with
a vacuum case. The dynamical equations become (again, we write
down only one $i$-th equation)

\begin{equation}
\begin{array}{l}
\sum\limits^D_{\substack{m=1 \\ m\ne i}} \left[ (p_m^2 - p_m)
\sum\limits_{\substack{\{j_1,\,j_2,\,\dots\,j_{2n-2}\}\ne\{i, m\}\\
j_1
> j_2 > \dots j_{2n-2}}}  p_{j_1} p_{j_2} \dots p_{j_{2n-2}}
\right] +
\\ \qquad + (2n-1)\sum\limits_{\substack{\{ k_1, k_2, \dots k_{2n}\}\ne i\\
k_1 > k_2 > \dots k_{2n}}} p_{k_1} p_{k_2} \dots p_{k_{2n}} = 0;
\end{array} \label{dyn_gen_p}
\end{equation}

\noindent and we omit the $t^{2p_m-2n}$ multiplier. Needless to
say, all the discussion in the previous section about the
difference between the cases with odd and even numbers of spatial
dimensions is valid for this {\it anzatz}, and the analogs of
%Eqs.\,(\ref{constr_odd, dyn_odd, constr_even, dyn_even}) could be
%Eqs.\,(10--13) could be easily written.
Eqs.\,(\ref{constr_odd}--\ref{dyn_even}) could be easily written.

Now let us find the conditions for the analogs of Kasner regime.
In finding them we followed~\cite{TT}. Let us consider the sum of
all dynamical equations -- the sum of all Eqs.\,(\ref{dyn_gen_p})
for all $i$. We will have a sum

\begin{equation}
\begin{array}{l}
\sum\limits_{[i]} \sum\limits^D_{\substack{m=1 \\ m\ne i}} \left[
(p_m^2 - p_m)
\sum\limits_{\substack{\{j_1,\,j_2,\,\dots\,j_{2n-2}\}\ne\{i, m\}\\
j_1
> j_2 > \dots j_{2n-2}}}  p_{j_1} p_{j_2} \dots p_{j_{2n-2}}
\right] +
\\ \qquad + (2n-1)\sum\limits_{[i]}\sum\limits_{\substack{\{ k_1, k_2, \dots k_{2n}\}\ne i\\
k_1 > k_2 > \dots k_{2n}}} p_{k_1} p_{k_2} \dots p_{k_{2n}} = 0.
\end{array} \label{kasner_p1}
\end{equation}

\noindent Taking (\ref{constr_gen_p}) into account, one can see
that the second term in (\ref{kasner_p1}) is zero. As for the
first term, it can be decomposed into two parts -- with only
linear multipliers and with only quadratic ones:

%\begin{equation}\label{kasner_p2}
$$
\begin{array}{l}
(D+1-2n) \sum\limits_{i=1}^{D}p_i^2 \sum\limits_{\substack{\{j_1,\,j_2,\,\dots\,j_{2n-2}\}\ne i\\
j_1 > j_2 > \dots j_{2n-2}}}  p_{j_1} p_{j_2} \dots p_{j_{2n-2}} - \\
- (2n-1)(D+1-2n) \sum\limits_{i > j_1 > j_2 > \dots j_{2n-2}} p_i
p_{j_1} p_{j_2} \dots p_{j_{2n-2}} = 0;
\end{array}
$$
%\end{equation}

\noindent the coefficients before the terms are the result of
simple combinatorics. Cancelling $(D+1-2n)$ one would obtain a
relation, which depends only on the order of Lovelock correction
and not on the number of spatial dimensions:

%\begin{equation}\label{kasner_p2}
%\begin{array}{l}
%\sum\limits_{i=1}^{D}p_i^2 \sum\limits_{\substack{\{j_1,\,j_2,\,\dots\,j_{2n-2}\}\ne i\\
%j_1 > j_2 > \dots j_{2n-2}}}  p_{j_1} p_{j_2} \dots p_{j_{2n-2}} + \\
%(2n-1)\sum\limits_{i=1}^{D}p_i \sum\limits_{\substack{\{j_1,\,j_2,\,\dots\,j_{2n-2}\}\ne i\\
%j_1 > j_2 > \dots j_{2n-2}}}  p_{j_1} p_{j_2} \dots p_{j_{2n-2}} =
%0;
%\end{array}
%\end{equation}

\begin{equation}\label{kasner_p2}
%\begin{array}{l}
\sum\limits_{i=1}^{D}p_i^2 \sum\limits_{\substack{\{j_1,\,j_2,\,\dots\,j_{2n-2}\}\ne i\\
j_1 > j_2 > \dots j_{2n-2}}}  p_{j_1} p_{j_2} \dots p_{j_{2n-2}} -
(2n-1) \sum\limits_{j_1 > j_2 > \dots j_{2n-1}}  p_{j_1} p_{j_2}
\dots p_{j_{2n-1}} = 0.
%\end{array}
\end{equation}

Now let's consider a sum

$$
\sum\limits_{i=1}^{D}p_i \sum\limits_{j_1 > j_2 > \dots j_{2n-1}}
p_{j_1} p_{j_2} \dots p_{j_{2n-1}}.
$$

\noindent This sum falls into the terms of two kinds -- if $i =
\{j_1, j_2, \dots j_{2n-1}\}$, its terms are like the first term
in (\ref{kasner_p2}), if $i \ne \{j_1, j_2, \dots j_{2n-1}\}$ --
like the second ones. After some simple combinatorics, one can
rewrite this sum as

$$
\begin{array}{l}
\sum\limits_{i=1}^{D}p_i \sum\limits_{j_1 > j_2 > \dots j_{2n-1}}
p_{j_1} p_{j_2} \dots p_{j_{2n-1}} = \sum\limits_{i=1}^{D}p_i^2
\sum\limits_{\substack{\{j_1,\,j_2,\,\dots\,j_{2n-2}\}\ne i\\
j_1 > j_2 > \dots j_{2n-2}}}  p_{j_1} p_{j_2} \dots p_{j_{2n-2}} +
\\ + 2n \sum\limits_{i > j_1 > j_2 > \dots j_{2n}} p_{i} p_{j_1} \dots
p_{j_{2n}}.
\end{array}
$$

\noindent From (\ref{constr_gen_p}) it is obvious that the second
term is equal to zero. So finally

\begin{equation}\label{kasner_p3}
\sum\limits_{i=1}^{D}p_i \sum\limits_{j_1 > j_2 > \dots j_{2n-1}}
p_{j_1} p_{j_2} \dots p_{j_{2n-1}} = \sum\limits_{i=1}^{D}p_i^2
\sum\limits_{\substack{\{j_1,\,j_2,\,\dots\,j_{2n-2}\}\ne i\\
j_1 > j_2 > \dots j_{2n-2}}}  p_{j_1} p_{j_2} \dots p_{j_{2n-2}}.
\end{equation}

Combining (\ref{kasner_p2}) and (\ref{kasner_p3}) results in

$$
\sum\limits_{i=1}^{D}p_i \sum\limits_{j_1 > j_2 > \dots j_{2n-1}}
p_{j_1} p_{j_2} \dots p_{j_{2n-1}} = (2n-1) \sum\limits_{j_1 > j_2
> \dots j_{2n-1}}  p_{j_1} p_{j_2} \dots p_{j_{2n-1}},
$$

\noindent so either

\begin{equation}\label{sum_p}
\sum\limits_{i=1}^{D}p_i = (2n-1),
\end{equation}

\noindent or

\begin{equation}\label{sum_3p}
\sum\limits_{j_1 > j_2 > \dots j_{2n-1}} p_{j_1} p_{j_2} \dots
p_{j_{2n-1}} = 0.
\end{equation}

First of the two is the condition for the analog of Kasner regime
in $(D+1)$-dimensional Universe with Lovelock gravity of the order
$n$, while the second corresponds to the  generalized Milne
solution (see the Discussion for more details).

\section{Influence of matter}

In this section we will investigate the influence the matter in
form of a perfect fluid exerts upon the dynamics of the system.
The equations of motion in this case take the form

\begin{equation}\label{eq_mot_mat}
G_{\mu\nu} = \frac{8\pi G_D}{3} T_{\mu\nu},
\end{equation}

\noindent where $G_D$ is a $D$-dimensional gravitation constant,
$T_{\mu\nu}$ is the stress-energy tensor of a perfect fluid

\begin{equation}\label{tmunu}
T_{\mu\nu} = (\rho + p) u_\mu u_\nu - p g_{\mu\nu} \quad
\mbox{with an equation of state}~p = w \rho,
\end{equation}

\noindent and $G_{\mu\nu}$ is a generalization of the Einstein
tensor in Lovelock gravity (Lovelock tensor; for issues about its
properties as a generalization of Einstein tensor, see, e.g.,
\cite{lovelock_tens}):

\begin{equation}\label{lovelock_tens}
G_{\mu\nu} = \frac{\delta{\cal L}}{\delta g^{\mu\nu}} -
\frac{1}{2}g_{\mu\nu}{\cal L} \quad \mbox{where}~{\cal L}~\mbox{is
the Lagrangian (\ref{lagr1}).}
\end{equation}

\noindent %One can easily see from (\ref{lovelock_tens}) that the
%way to obtain the Lovelock tensor is the same as obtaining the
%Einstein tensor. So we can call Lovelock tensor a generalization
%of Einstein's one and claim the validity of (\ref{eq_mot_mat}).
So, the equations of motion would take a form: \\for constraint

\begin{equation}
(2n-1) \sum\limits_{k_1 > k_2 > \dots k_{2n}} H_{k_1} H_{k_2}
\dots H_{k_{2n}} = \rho; \label{constr_mat}
\end{equation}

\noindent and for dynamical equations (again, $i$-th equation)

\begin{equation}
\begin{array}{l}
\sum\limits^D_{\substack{m=1 \\ m\ne i}} \left[ (\dot H_m + H_m^2)
\sum\limits_{\substack{\{j_1,\,j_2,\,\dots\,j_{2n-2}\}\ne\{i, m\}\\
j_1
> j_2 > \dots j_{2n-2}}}  H_{j_1} H_{j_2} \dots H_{j_{2n-2}}
\right] +
\\ \qquad + (2n-1)\sum\limits_{\substack{\{ k_1, k_2, \dots k_{2n}\}\ne i\\
k_1 > k_2 > \dots k_{2n}}} H_{k_1} H_{k_2} \dots H_{k_{2n}} + p =
0.
\end{array} \label{dyn_mat}
\end{equation}

\noindent In the equations (\ref{constr_mat}, \ref{dyn_mat}), we
renormalized the density $\rho$ in order to avoid the $8\pi G_D
/3$ term -- we are interested in the qualitative description so we
can do this. And of course, in addition to (\ref{constr_mat},
\ref{dyn_mat}) one needs a continuity equation:

\begin{equation}\label{cont_eq}
\dot \rho + (\rho + p) \sum\limits_{i=1}^D H_i = 0.
\end{equation}

The system (\ref{constr_mat}--\ref{cont_eq}), along with initial
conditions, completely determines the dynamics of the system and
its past and future evolution.

Let us rewrite (\ref{constr_mat}, \ref{dyn_mat}) for the power-law
{\it anzatz}. It's easy to see that Lovelock tensor scales as

\begin{equation}\label{lovelock_scale}
\begin{array}{l}
G_{00} = (2n-1) t^{-2n} \sum_1 p_i, \\
G_{jj} = t^{2p_j - 2n} \sum_2 p_i,
\end{array}
\end{equation}

\noindent where $\sum_1 p_i$ and $\sum_2 p_i$ are the
corresponding sums in l.h.s. of Eqs.\,(\ref{constr_gen_p}) and
(\ref{dyn_gen_p}), respectively (one can easily verify that both
terms in Eq.\,(\ref{dyn_gen_p}) scale the same). Solving
(\ref{cont_eq}) in the chosen {\it anzatz}, one can write down the
scaling of the stress-energy tensor components:

\begin{equation}\label{tmunu_scale}
\begin{array}{l}
T_{00} = \rho_0 t^{-(1+w) \sum p_i}, \\
T_{jj} = w \rho_0 t^{2p_j - (1+w)\sum p_i}.
\end{array}
\end{equation}

\noindent Matching (\ref{lovelock_scale}) and (\ref{tmunu_scale}),
one can find the condition under which system can be solved:

\begin{equation}\label{sum_p_mat}
\sum\limits_{i=1}^{D}p_i = \frac{2n}{1+w}.
\end{equation}

Finally, matching (\ref{sum_p}) with (\ref{sum_p_mat}), one can
find the ``critical'' value for the equation of state, with
separate Lovelock and matter-dominating regimes (see the
Discussion for more details):

\begin{equation}\label{w_cr}
w_{cr} = \frac{1}{2n -1}.
\end{equation}

\section{Discussion}

We have derived the vacuum equations of motion (\ref{constr_gen},
\ref{dyn_gen}) for Bianchi-I-type metric for any number of spatial
dimensions ($D\geqslant 4$) in the Lovelock gravity of any order
$n\geqslant 2$. The corresponding Lagrangian for $L_2$ was first
derived by Lanczos~\cite{Lanczos} and for $L_3$ in~\cite{d4}. One
can verify, that for the chosen metric our result coincides with
the previously obtained ones.

If one considers only the leading (in $1/t$ order) terms, then one
needs to take into consideration only the highest-order Lovelock
contribution, it would be $n=\[D/2\]$. Therefore we rewrote our
equations of motion for the cases of odd $D=2n+1$
(\ref{constr_odd}, \ref{dyn_odd}) and even $D=2n$
(\ref{constr_even}, \ref{dyn_even}) numbers of spatial dimensions.
One can easily see the difference between the two: while there are
no restrictions from constraint on Hubble functions in case of odd
numbers [of spatial dimensions] (Eq.\,(\ref{constr_odd})), the
even-numbers case (Eq.\,(\ref{constr_even})) sets a restriction:
one of Hubble parameters should be always equal to zero. The
dynamical equations also differ in structure ((\ref{dyn_odd}) vs.
(\ref{dyn_even})). Our numerical investigation of (4+1) and (5+1)
cases in $L_2$~\cite{new} proved that there is a significant
difference between two cases. We believe the difference in $L_2$
is due to the mentioned above reasons, so one should expect
similar features from higher-dimension cases as well.

When it comes to the anisotropic cosmological models in Lovelock
gravity, two main cases are considered: product spaces and
Bianchi-I models. As we claimed in Introduction, we consider the
latter one. And for Bianchi-I-type models most of results are
obtained in power-law {\it anzatz}. We also applied this {\it
anzatz} to our solution and investigated it a bit. It has a form
(\ref{constr_gen_p}, \ref{dyn_gen_p}) and coincides with
previously found results for corresponding $n$ (in addition to the
previously mentioned papers, let us add~\cite{9d} with a result
for [$D=9$, $L_4$]).
%This again proves that our result is correct.% (well, at least up
%to $L_4$).

For the power-law {\it anzatz} we also found an expansion rate in
terms of Kasner exponents: $\sum_i p_i = (2n-1)$. This result also
coincides with those previously reported for low $n$. Finally, we
found the condition on $p_i$, which corresponds to the generalized
Milne solution. It is determined from two conditions --
(\ref{constr_gen_p}) and (\ref{sum_3p}). And here we can again see
the difference between the cases with odd and even numbers of
spatial dimensions. Indeed, for $D=2n$, the
Eq.\,(\ref{constr_gen_p}) reduces to only one term directly
implying that one of $p_i$ should be zero. With this,
Eq.\,(\ref{sum_3p}) also reduces to only one term implying another
$p_i$ should also be zero; the rest of $p_i$ are arbitrary. Thus,
the resulting $p_i$ layout is as follows: ($a_1$, $a_2$, \dots,
$a_{D-2}$, 0, 0). For $D=2n+1$, the situation is a bit different
-- since $D=2n+1$, we have $D$ terms in Eq.\,(\ref{constr_gen_p}),
and nullifying one of $p_i$ will leave us with only one term in
l.h.s. of Eq.\,(\ref{constr_gen_p}), and this eventually nullifies
another $p_i$. But now with two of $p_i$ set to zero, the l.h.s.
of Eq.\,(\ref{sum_3p}) is also left with only one term and this
sets third of $p_i$ to zero. Therefore for odd numbers of spatial
dimensions the resulting $p_i$ layout is as follows: ($a_1$,
$a_2$, \dots, $a_{D-3}$, 0, 0, 0). Finally, in a $D$-independent
form: ($a_1$, $a_2$, \dots, $a_{2n-2}$, 0, \dots). A proof that
there are no solutions except the two mentioned above (Kasner-like
with (\ref{sum_p}) and generalized Milne with (\ref{sum_3p})), is
quite similar to the proof for $L^5_2$~\cite{mpla}.

Finally we studied possible influence of matter in the form of
perfect fluid on the dynamics of the system. We wrote the
equations of motion (\ref{constr_mat}--\ref{cont_eq}) and deepened
the difference between the cases with odd and even numbers of
spatial dimensions. In the presence of matter, even-dimensional
cases change drastically -- without the matter one of Hubble
functions should be zero (see Eq.\,(\ref{constr_even})), while
even a small amount of matter changes the situation -- in the
presence of matter ($\rho\ne 0$) none of Hubble parameters can be
zero (see Eq.\,(\ref{constr_mat})).

We as well found a condition for the system to be solvable in the
power-law {\it anzatz} (Eq.\,(\ref{sum_p_mat})). Matching
(\ref{sum_p}) and (\ref{sum_p_mat}), one can find a ``critical''
value for the equation of state, with separates Lovelock and
matter-dominating regimes. In~\cite{new} we demonstrated that this
value separates two different asymptotic [leading to initial
singularity] regimes -- if $w > w_{cr}$ the initial singularity is
isotropic (and the matter terms are dominating), and if $w <
w_{cr}$ the initial singularity is anisotropic (and the
Gauss-Bonnet terms are dominating). We believe that the same
situation holds in the higher orders of Lovelock gravity, though
some numerical simulations might be an asset. Finally, if one
rewrite Eqs. (\ref{constr_mat}) and (\ref{dyn_mat}) for $D=2n$ and
$w=0$ in terms of $p_i$, they would become $p_{k_1} p_{k_2} \dots
p_{k_{2n}} = \rho$ for the constraint and

\begin{equation}\nonumber
\sum\limits^D_{\substack{m=1 \\ m\ne i}}  p_m (p_m - 1) ~p_{j_1}
p_{j_2} \dots p_{j_{2n-2}} = 0
\end{equation}

\noindent for the dynamic equations (there we omit the powers of
$t$). One can see, that the ``special solution'' $p_i \equiv 1$,
which was described in~\cite{new} for $L^4_2$, is valid for the
general case of Lovelock gravity in even numbers of spatial
dimensions.

%\begin{equation}
%\begin{array}{l}
%p_{k_1} p_{k_2} \dots  p_{k_{2n}} = \rho \quad \mbox{for
%constraint and} \nonumber \\
%\sum\limits^D_{\substack{m=1 \\ m\ne i}}  p_m (p_m - 1) ~p_{j_1}
%p_{j_2} \dots p_{j_{2n-2}} = 0 \quad \mbox{and we omitted powers
%of}~t.
%\end{array}
%\end{equation}

And the last, but nonetheless a very important note. All through
the present paper we claim that our results are obtained with only
one highest-order Lovelock correction taken into account. In
reality, some of them can be generalized for a mixture of Lovelock
corrections with different $n$. If one wants to deal with some
different Lovelock corrections at once (this would be the case if
one wants to study the dynamics not only in the vicinity of an
initial singularity), one should use (\ref{lagr}) instead of
(\ref{lagr1}) as a Lagrangian density and substitute all
individual $L_n$ needed into it. But in this case one need to use
(\ref{big_lagr1}) expression for $L_n$ since we cancel $2^{-n}
\times (2n-3)!!$ terms in (\ref{big_lagr2}) in order to simplify
it for the case where only one $n$ is used. Now if we combine this
$n$-dependent multiplier with $\alpha_n$ into new constant
$\tilde\alpha_n$, equations take a form

%\noindent where $\alpha_i$ is the length scale corresponding to
%$L_i$, and the summation comes over all the needed $n$ (it could
%be from 1 to the highest possible $n$ if one wants to take all
%terms into account or just some specific $n$ for specific
%purposes). The equations of motion derived from~(\ref{lagr_gen})
%take a form of

\begin{equation}\label{eqmot_gen}
\sum_{i}\tilde\alpha_i P_i = 0,
\end{equation}

%\noindent where $\tilde\alpha_i$ is the multiplier which takes
%into account $\alpha_i$ from (\ref{lagr_gen}) and multipliers
%omitted when we simplified (\ref{big_lagr1}) into
%(\ref{big_lagr2});
\noindent where $P_i$ are the equations of motion corresponding to
$L_i$ -- l.h.s. of Eq.\,(\ref{constr_gen}) for a constraint or
Eq.\,(\ref{dyn_gen}) for a dynamical equation. Thus, equations can
be easily constructed from the already derived ones. The same
comes for the matter-filled universe, but in this case in the
r.h.s. of (\ref{eqmot_gen}) would be $\rho$ for a constraint or
$(-p)$ for the dynamical equations (compare with
(\ref{constr_mat}) and (\ref{dyn_mat})).

From the way we derived the equations for the power-law {\it
anzatz} one can see that the power-law analogs of equations cannot
be generalized for the mixture of $n$. This comes out from the way
we derived them -- as you remember, we omitted the powers of $t$
since they were the only ones for vacuum case or the same with
those in r.h.s. for the matter case. But with a mixture of $n$ we
would have different powers of $t$ for different $n$ and so we
cannot cancel them anymore. With this, the results obtained from
the power-law {\it anzatz} (like different sums of $p_i$), cannot
be applied for the case with mixture of $n$. Yet, the results for
the highest possible $n$ can be used as an asymptotic regime for
this case.

%\section{Conclusions}

%We have studied many aspects of Bianchi-I cosmological models in
%Lovelock gravity. One of our main achievements -- we wrote down
%Lagrangian and equations of motion for vacuum models and those
%filled with matter in form of perfect fluid. Our equations are
%generic for any order of Lovelock correction and any number of
%spatial dimensions, which make them very useful for further study
%of different aspects of Bianchi-I cosmological models. Obtained
%equations can also be very easily modified for different kinds of
%stress-energy tensor, which make them even more useful.

\section{Conclusions}

In this paper we investigated Bianchi-I cosmological models in
Lovelock gravity. We were first to derive the Lagrangian and the
equations of motion for the model considered: Bianchi-I model in
Lovelock gravity with only highest possible correction taken into
account. This by itself is a great achievement -- previously, when
one deals with Bianchi-I models in Lovelock gravity, only one
specific order was involved. This is linked with a difficulty in
obtaining equations from the most general case. In our paper we
imply symmetries of Bianchi-I model to obtain our results; our
result is correct only for Bianci-I case but it cover all possible
$n$ as well as is valid for any number of spatial dimensions.

We analyzed obtained equations of motion separately for the cases
of odd and even number of spatial dimensions and proved that there
is a difference between these two cases.

After applying power-law {\it anzatz} to our equations of motion
we were able to obtain conditions for the Kasner as well as
generalized Milne regimes. Results obtained matches those for
low-$n$ cases and generalize them.

Finally, we took the matter in form of perfect fluid into
consideration. For this case we also obtained equations of motion;
condition this equations can be resolved under and the
``critical'' value for the equation of state which separates
Lovelock-dominating from matter-dominating regimes. Results of the
analysis of the model with matter make the mentioned above
difference between $D=2n$ and $D=2n+1$ cases even deeper.

We also discussed the way of constructing the equations of motion
for a mixture of $n$ from the obtained equations both for vacuum
and for matter cases.

All results obtained matches with existing results for low-$n$
cases and generalize them for all possible $n$.

\acknowledgements

I would like to thank A.V. Toporensky and I.V. Kirnos for a
stimulating discussion and S.V. Karpov for technical help.

\end{document}